\documentclass[10pt,letterpaper]{article}
\usepackage{opex3}
\usepackage{hangcaption}
\usepackage{array}
\usepackage{indentfirst} 
\usepackage{cite}

\begin{document}




\title{Random phase-free kinoform \\
for large objects}

\author{
Tomoyoshi Shimobaba,$^{1*}$
Takashi Kakue,$^1$
Yutaka Endo,$^{1}$\\
Ryuji Hirayama,$^{1}$
Daisuke Hiyama,$^{1}$
Satoki Hasegawa,$^{1}$\\
Yuki Nagahama,$^{1}$
Marie Sano,$^{1}$
Takashige Sugie,$^{1}$
and 
Tomoyoshi Ito$^1$
}

\address{$^1$ Graduate School of Engineering, Chiba University, 1-33 Yayoi-cho, Inage-ku, Chiba 263-8522, Japan}

\begin{abstract} 
We propose a random phase-free kinoform for large objects.
When not using the random phase in kinoform calculation, the reconstructed images from the kinoform are heavy degraded, like edge-only preserved images.
In addition, the kinoform cannot record an entire object that exceeds the kinoform size because the object light does not widely spread.
In order to avoid this degradation and to widely spread the object light, the random phase is applied to the kinoform calculation; however, the reconstructed image is contaminated by speckle noise.
In this paper, we overcome this problem by using our random phase-free method and error diffusion method.
\end{abstract}


\


\section{Introduction}
\noindent Computer-generated holograms (CGHs)  \cite{poon} are important in optics and are used for three-dimensional television, diffractive optical elements, projection and encryption.
CGHs are mainly categorized into three types: complex amplitude CGH, amplitude CGH and kinoform \cite{kinoform}, which is also known as phase-only CGH.
Complex amplitude CGH is capable of reconstructing a perfect image; however, complex amplitude CGH requires a special device that displays  the real and imaginary parts (otherwise, amplitude and argument parts) of the complex amplitude.
Whereas, amplitude CGH and kinoform do not need such a special device.
Amplitude CGH or kinoform is calculated by taking only the real part or the argument part of object light.
The reconstructed image quality is degraded, compared to complex amplitude CGH.
This paper focuses on kinoform.

Kinoform can reconstruct bright images, compared to amplitude CGHs, because the kinoform has theoretically 100 \% light efficiency.
In addition, reconstructed images from kinoforms do not include conjugate light, unlike amplitude CGHs. 
In kinoform calculation, applying the random phase to objects for widely diffusing the object light is necessary to record the entire object on the kinoform.
When not using the random phase, the reconstructed images from the kinoform are heavy degraded, like edge-only preserved images.
In addition, the kinoform cannot record an entire object that exceeds the kinoform size because the object light does not widely spread.
In order to avoid this degradation and to widely spread the object light, the random phase is applied to the kinoform calculation.
Although the problem of edge-only images and the narrow spreading object light can be improved by the random phase, the reconstructed image is contaminated by speckle noise.

The well-known methods for improving the speckle noise are the iterative methods \cite{ite1,ite2} such as Gerchberg-Saxton (GS) algorithm, multi-random phase method \cite{amako}, one step phase retrieval method (OSPR) \cite{ospr}, and pixel separation methods \cite{pix1,pix2,pix3}.
These methods improve the speckle noise; however, they are time-consuming because the multiple diffraction calculations and plural CGHs are needed. 
Additionally, the multi-random phase method and pixel separation methods require display devices with high-speed refresh rates.

On the other hand, random-phase free methods without time-consuming processing have been proposed: for example, the error diffusion methods \cite{ospr,ed1,ed2,ed3,ed4} and down-sampling method \cite{down}.
These methods can reconstruct clear images, not edge-only preserved images; however, the size of the reconstructed image cannot exceed the kinoform size because the object light does not widely spread.
Therefore, these methods cannot be used for large objects that are required in the applications of holographic projection and three-dimensional display \cite{zoom1,zoom2,zoom3,zoom4}.
Recently, a new random phase-free method using a virtual special convergence light has been proposed \cite{random_free}. 
This method can reconstruct large images that exceed the CGH size with low speckle noise. 
This method is powerful; however, the effectiveness of the method is only shown in amplitude CGH. 

In this paper, we propose random phase-free kinoform for large objects by the combination of our random phase-free method \cite{random_free} and error diffusion methods.
Kinoforms generated by the proposed method can reconstruct large images, which exceed the CGH size, with low speckle noise. 

\section{Proposed method}
\begin{figure}
\centerline{\includegraphics[width=13cm]{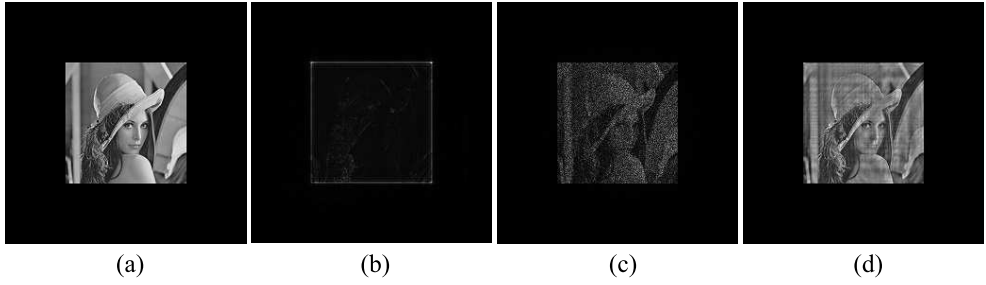}}
\caption{Original and reconstructed images. (a) original image (b) without random phase (c) with random phase (d) with error diffusion method}
\label{fig:reconst-x1}
\end{figure}

\noindent  To begin with, in Fig.\ref{fig:reconst-x1}, we show the reconstructed images from kinoforms without random phase, with random phase, and with the error diffusion method proposed by Ref. \cite{ed3}, respectively.
In this simulation, we use the wavelength of 532 nm, the pixel pitch of 8 $\mu$m, the resolution of 2,048 $\times$ 2,048 pixels, and the propagation distance of 0.8 m between the kinoform and the image. 

\begin{figure}
\centerline{\includegraphics[width=13cm]{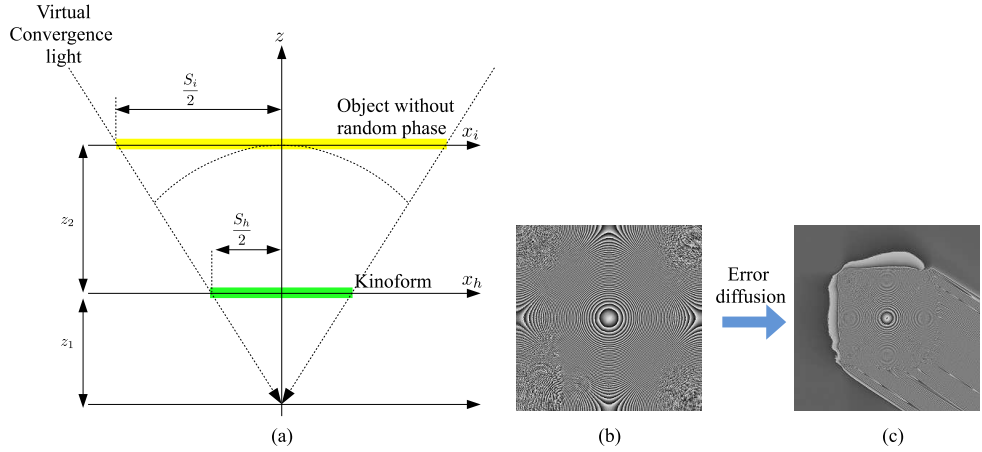}}
\caption{Proposed method. (a) calculation setup for random phase-free kinoform (b) generated kinoform (c) error diffused kinoform.}
\label{fig:propose}
\end{figure}

In. Fig.\ref{fig:reconst-x1}(a) shows the original image and (b) shows the reconstructed image from a kinoform without random phase.
The reconstructed image preserves only the edge information \cite{ed3,ed4,down}. 
Fig.\ref{fig:reconst-x1}(c) shows the reconstructed image from a kinoform with random phase, and we can recognize the original information; however, the image is contaminated by speckle noise.
Fig.\ref{fig:reconst-x1}(d) shows the reconstructed image from a kinoform with the error diffusion method \cite{ed3}, and the image has the best image quality in these reconstructed images.
Unfortunately, as will be shown in Section 3, these methods cannot be used in a situation in which the original image size exceeds the kinoform.  

In this section, we describe how to overcome this problem by the proposed method based on our random phase-free method \cite{random_free}. 
As shown in Fig.\ref{fig:propose}, this method applies virtual spherical  convergence light to the input image instead of random phase.
The areas of the image and the kinoform are $S_{i} \times S_{i}$ ($S_i =N \times p_i$) and $S_{o} \times S_{o}$ ($S_o =N \times p_o$), where $N$ is the number of pixels for each side of the image, $p_i$ and $p_o$ are the sampling pitches on the object and kinoform, respectively.
In the proposed method, we multiply the image $u_i(x_i, y_i)$ by the convergence light given by $w(x_i, y_i)$.
 $w(x_i, y_i)$ is expressed as a convergence spherical light,
\begin{equation}
w(x_i, y_i)=\exp(-i \pi(x_i^2+y_i^2)/\lambda f_i)
\end{equation}
where $f_i=z_1+z_2$ is the focal length.
The distance between the object and the kinoform is denoted by $z_2$.
The distance between the focus point of the convergence light and the kinoform is denoted by $z_1$, and is set to the distance at which the kinoform just fits to the cone of the convergence light.
Subsequently, we calculate the complex amplitude on the kinoform using,
\begin{equation}
u_h(x_h, y_h) = {\rm Prop_{z_2}}\{u_i(x_i, y_i) w(x_i, y_i)\}, 
\end{equation}
where $ {\rm Prop_{z_2}\{\cdot \}}$ denotes the diffraction calculation at the propagation distance $z_2$.
Using a simple geometric relation as shown in Fig.\ref{fig:propose}, we can derive,
\begin{equation}
S_{h}/2 : S_{i}/2 = z_1: f_i, 
\end{equation}
hence $f_i=z_2/(1-S_{h}/ S_{i})$.
See more details in Ref. \cite{random_free}.
Finally, we obtain the kinoform by,
\begin{equation}
\theta(x_h, y_h) = {\rm arg} \{u_h(x_h, y_h) \} 
\end{equation}
where ${\rm arg} \{\cdot \} $ indicate the use of only the argument part of complex amplitude. 
The generated kinoform is shown in Fig.\ref{fig:propose}(b).
Our random phase-free method worked well in amplitude CGH; however, in kinoform, our random phase-free method reconstructs an edge-only preserved image like Fig. \ref{fig:reconst-x1}(b).
To overcome this problem, we apply the error diffusion method \cite{ed3} to the kinoform.
The error diffusion method we used calculates,  
\begin{eqnarray}
\theta(x_h, y_h+1) &\leftarrow&  \theta(x_h, y_h+1) + w_1 e(x_h, y_h), \\
\theta(x_h+1, y_h-1) &\leftarrow&  \theta(x_h+1, y_h-1) + w_2 e(x_h, y_h), \\
\theta(x_h+1, y_h) &\leftarrow&  \theta(x_h+1, y_h)  + w_3 e(x_h, y_h), \\
\theta(x_h+1, y_h+1) &\leftarrow&  \theta(x_h+1, y_h+1) + w_4 e(x_h, y_h), 
\end{eqnarray}
where we use Floyd-Steinberg coefficients $w_1$=7/16, $w_2$=3/16, $w_3$=5/16, and $w_4$=1/16 \cite{ed}, $\leftarrow$ indicates the current value is overwritten, and $e(x_h, y_h)=u_h(x_h, y_h) -\theta(x_h, y_h)$.
Before applying the error diffusion, we may need to normalize the complex amplitude $u_h(x_h, y_h)$,
\begin{equation}
u_h(x_h, y_h) \leftarrow u_h(x_h, y_h) / d,
\end{equation}
where $d={\rm max}|u_h(x_h, y_h)|$ is the maximum absolute value in the kinoform. 
The error diffused kinoform is shown in Fig.\ref{fig:propose}(c).

\section{Results}
\begin{figure}
\centerline{\includegraphics[width=12cm]{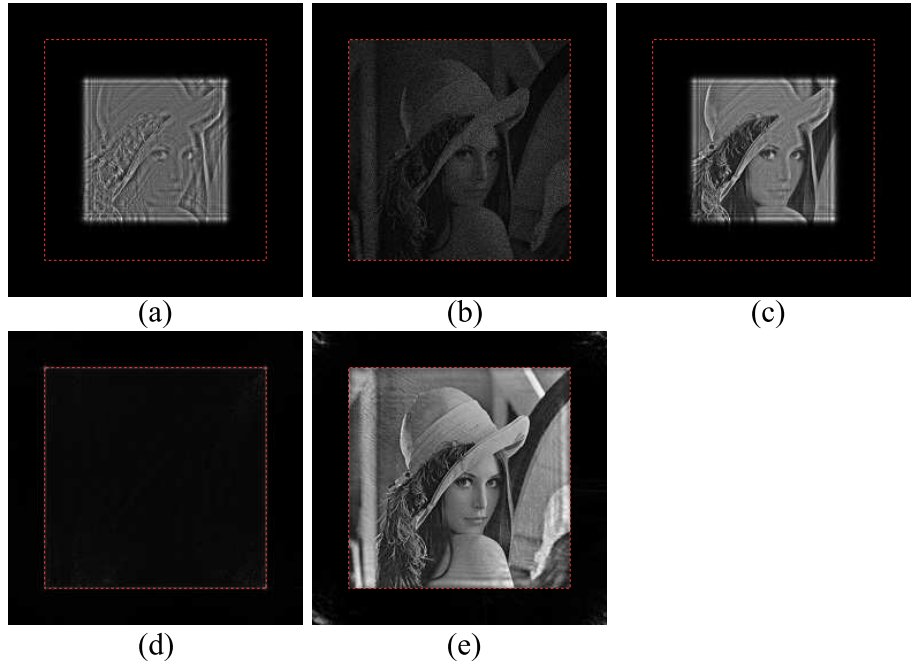}}
\caption{Reconstructed images. (a) without random phase (b) with random phase (c) with error diffusion (d) with our random phase-free method (e) with our random phase-free method and error diffusion. The red dashed boxes indicate the correct image size.}
\label{fig:reconst-x3}
\end{figure}

\noindent Figure \ref{fig:reconst-x3} shows the reconstructed images from kinoforms without random phase (Fig.\ref{fig:reconst-x3}(a)), with random phase  (Fig.\ref{fig:reconst-x3}(b)),  with error diffusion (Fig. \ref{fig:reconst-x3}(c)), with our random phase-free method  (Fig. \ref{fig:reconst-x3}(d)), and with our random phase-free method and error diffusion (Fig. \ref{fig:reconst-x3}(e)) .
The calculation conditions are the same as those shown in Fig. \ref{fig:reconst-x1}, except for the sampling pitch on the original image of $p_i=24 \mu$m.
The sampling pitch of the kinoform is $p_h=8 \mu$m.
We used scaled diffraction \cite{arss} to calculate the diffraction calculation from the object to the kinoform at different sampling pitches.
The size of the reconstructed images is expected to be three times (=$p_i/p_h$) larger than the kinoform size.
Therefore, the kinoform size is 16.4 mm $\times$ 16.4 mm ($\approx 2,048 \times 8 \mu$ m $\times $$2,048 \times 8 \mu$ m), and the sizes of the original and reconstructed images are 49.2 mm $\times$ 49.2 mm ($\approx  2,048 \times 24 \mu$ m $\times $$2,048 \times 24 \mu$ m).
The red dashed boxes indicate the correct image size.

In Figs.\ref{fig:reconst-x3}(a) and (c), these reconstructed images do not have the correct image size because the light from the original image does not spread over the kinoform during the kinoform calculation.
In Figs.\ref{fig:reconst-x3}(b), the reconstructed image has the correct image size thanks to the random phase; however, the speckle noise occurs.
In Figs.\ref{fig:reconst-x3}(d), the reconstructed image has the correct image size; however, it becomes an edge-only preserved image.
Whereas, the reconstructed image by our random phase-free method and error diffusion method (Fig.\ref{fig:reconst-x3}(e)) has better image quality and correct image size.
The peak signal-to-noise ratios (PSNRs) between the original image and these reconstructed images are 10 dB (a), 12 dB(b), 10 dB (c), 6 dB (d), and 22 dB (e), respectively.
All of the results in this paper were obtained by our numerical library for wave optics \cite{cwo}.

\section{Conclusion}
\noindent We proposed random phase-free kinoform for large objects using our random phase-free method \cite{random_free} and error diffusion \cite{ed3}.
The kinoform calculation for large objects is useful for holographic projection \cite{zoom1} and holographic three-dimensional display \cite{zoom4}.
The random phase, which is main generating factor of speckle noise, is necessary in the kinoform calculation of large objects, for reconstructing the correct size.
The proposed method is capable of reconstructing the image with low speckle noise, and is computationally inexpensive.

\section*{Acknowledgments}
\noindent This work is partially supported by JSPS KAKENHI Grant Numbers 25330125 and 25240015, and the Kayamori Foundation of Information Science Advancement and Yazaki Memorial Foundation for Science and Technology.

\end{document}